# Mechanistic Modelling of Chromatin Folding to Understand Function

Chris A. Brackley[1], Davide Marenduzzo[1], and Nick Gilbert[2]

[1]*SUPA School of Physics and Astronomy, University of Edinburgh, Edinburgh EH9 3FD, UK.*

[2]*MRC Human Genetics Unit, Institute of Genetics & Molecular Medicine, University of Edinburgh, Edinburgh, UK.*

Correspondence to:

Nick.Gilbert@ed.ac.uk or Davide.Marenduzzo@ed.ac.uk

**Abstract**

**Experimental approaches have been applied to address questions in understanding three-dimensional chromatin organisation and function. As datasets increase in size and complexity, it becomes a challenge to reach a mechanistic interpretation of experimental results. Polymer simulations and mechanistic modelling have been applied to explain experimental observations, and the links to different aspects of genome function. Here, we provide a guide for biologists, explaining different simulation approaches and the contexts in which they have been used.**

**Introduction**

Technological advances in next-generation sequencing and microscopy have led to an explosion of new methods to probe genome structure and function. As the size and complexity of experimental datasets increase, it is not always easy to understand and interpret the underlying biology. Chromosome-conformation-capture based methods, such as Hi-C, (see reference 1 for a review paper discussing 3C-based techniques[1]) provide a population level readout of how likely two genomic loci are to be in spatial proximity and how genome interactions are structured. At large scales, interactions between different active chromatin regions (and interactions between different inactive chromatin regions) are enriched compared to active-inactive interactions. In this way, there is a segregation of the genome into compartments[2]. At shorter length scales, chromosomes are partitioned into contiguous regions which show enriched self-interactions, known as domains or TADs (topologically associating domains)[3]; many domains are associated with chromatin loops (between, for example, binding sites of the CCCTC-binding factor, CTCF)[4]. More recently, higher resolution data sets have revealed features such as "stripes" of enriched interaction emanating from super enhancers[5,6] (also called "frequently interacting regions" or FIREs). As experiments have become more quantitative, researchers are turning to methods commonly used in other data-rich fields: for example, statistical modelling, machine learning, and – the subject of this perspective – mechanistic modelling. Mechanistic models, which seek to understand *why* a system behaves as it does, are common in the physical sciences; they are often based on simplified mathematical or computational representations of a system, fitting together individual components to study emergent behaviour.

Here we focus on mechanistic models which originate from soft matter and polymer physics. Soft matter physics is the study of "soft" materials such as gels, colloids and polymers, the properties of which are often found in biology – in tissues, the cytoplasm, and the nucleus. The defining feature of soft matter systems is that their macroscopic behaviour depends on their mesoscopic properties: behaviour is governed by thermal fluctuations, self-organisation and entropy. That is to say, the properties of these systems are often not determined by microscopic or molecular details, which permits simplified "coarse-grained" approaches for modelling these systems[7]. While molecular details are of course crucial for biological functions, the mesoscale approach has often proved fruitful[8].

In this perspective we aim to demystify the concepts of mechanistic modelling in the context of chromatin folding, chart progress to date, and provide a view on what the future holds.

**The approach to modelling: inverse or forwards.**

When modelling the three-dimensional (3D) structure of chromosomes, there are two commonly used approaches: inverse (or "fitting") based models and forward (or "mechanistic") models. In inverse modelling, one starts with experimental data, such as chromosome interactions e.g. from 5C or Hi-C, and uses that information to reconstruct the spatial arrangement of the chromosome in 3-D, via a fitting or an iterative procedure. Early attempts used the data to infer a set of constraints on the separation of different genomic loci, using those as a basis for statistical or matrix methods to generate a single "average" chromosome conformation[9]. More recently, methods from polymer physics – where the chromatin fibre is represented as a connected chain of interacting units – have been applied[10,11]; the interaction parameters are then determined using an iterative procedure which improves the fit of the model. Usually, the polymer approaches generate a population of 3-D conformations, and in some cases once an optimal parameter set is found, the same model can be applied to different chromosomes or cell types[11]. Inverse modelling can be a useful tool – especially when there is a lot of existing data. For example, early studies revealed the importance of loops[12], and that the observed interaction maps are consistent with a population of chromosome structure which has a high degree of variability[10].

Though inverse modelling can successfully reproduce, for example, Hi-C interaction maps, it often reveals little about the mechanisms behind the observed structures. In mechanistic (or forward) modelling the aim is different. Instead of reproducing experimental data, mechanistic modelling offers a way to understand the mechanisms behind the observations, e.g. addressing questions such as how does a chromosome segregate into active and inactive compartments, or how do chromatin domains form? Mechanistic models usually start from first principles or hypotheses and ask whether different microscopic possibilities can generate the observed behaviours. A common approach (which is prevalent in the physical sciences) is to start with a simple description of the system (i.e. a minimal model), and then systematically add detail until the model reproduces the experimental observations. Such models can also incorporate experimental data, but the "output" is usually a different kind of information from the "input". While in the inverse modelling case the input might be a Hi-C map, and the output will be a reproduced Hi-C map along with a 3-D structure. In contrast a mechanistic model might use protein binding data together with some hypothesised "rules" to generate 3-D structures and a simulated Hi-C map (but, Hi-C is *not* used as an input). In this way mechanistic modelling is truly predictive, but the accuracy of the prediction is unlikely to be as good as a fitting based model with many (or hundreds) of parameters; nevertheless, they have been hugely informative and provide an unique understanding of molecular mechanisms.

**Dynamical Polymer Models: Key Concepts and Ideas**

Most mechanistic models of chromatin are derived from polymer physics and molecular dynamics (MD) simulations[13]. MD is a scheme where the motions of atoms or molecules are simulated in silico. The principle is simple: atoms move according to Newton's second law, which states that the force exerted on a body is equal to its mass multiplied by its acceleration (Figure 1A). This equation of motion is solved to simulate how the positions of all the atoms in a system change with time. After deciding where to initially place the atoms, the remaining problem is to calculate the forces which each atom will experience. In practice this can be challenging: each atom will experience forces due to interactions with all of the surrounding atoms (Coulomb interactions due to charge, van der Waals interactions etc.); nevertheless, it has been possible to write down a set of interaction potentials (based on calculations from quantum chemistry, collectively described as a "force field") which results in realistic behaviour. Such "all-atom" MD simulations have been applied to proteins, as well as DNA and DNA-protein interactions. However, calculating the forces (including those between solvent atoms such as water and salt ions etc.) is computationally expensive, and typically this can only be done for a short DNA segment or a single protein subunit (for example it currently takes around a day to simulate the behaviour of a protein consisting of 142,000 atoms for 100 ns on 256 compute cores, or about 0.3 ns per day per core).

To simulate larger experimental systems, one must adapt the method – one way to do this is by removing the solvent from the simulation. Instead of explicitly treating all water molecules, the effects can be included in an implicit way. Common methods include Brownian or Langevin Dynamics; the simplest method is to include only the viscous drag and "thermal jostling" which water provides. This leads to an accurate representation of the thermal diffusion properties. (Figure 1Bi and see Box 1 for details). More complicated approaches might include hydrodynamic interactions or include the effect of salt ions via further modification of the interaction force field, e.g. to account for electrostatic screening.

Even after removing the solvent, the size of systems which can be simulated with an all-atom treatment remains relative small. Also, keeping track of the position of every atom within a chromosome is unlikely to be informative. To circumvent these problems, large systems are often coarse-grained (CG): collections of atoms or molecules are replaced by larger and simpler objects, meanwhile the same method of solving Newton's law with an implicit solvent can be used (Figure 1Bii-iii). It is also necessary to have a scheme for determining the forces that the simplified objects will exert on each other – these interaction potentials are often phenomenological in nature (i.e. one uses a set of interactions which provide correct macroscopic behaviour but which do not need to be microscopically realistic). Yet, it is not an easy task to develop a CG model. The difficulty lies in deciding on how much detail to include in a model to learn something meaningful. This of course depends on the question (see Table 1): to investigate how a polymerase bends and twists DNA during transcription it might be important to model the double helix of the DNA[14,17]; on the other hand, to understand how a chromatin fibre is formed, perhaps the double helix structure is less important, but a description of the nucleosome needs to be included[18]. To model a whole chromosome or chromosome region, most work has used a much lower level of detail where a chromatin fibre is represented by a connected chain of beads, with each bead representing several thousand base-pairs of DNA. Such large-scale simulations have been informative for understanding chromatin organisation.

While a chain of beads might sound like a crude model for chromatin, it has been remarkably successful in explaining many of the observations from experiments like Hi-C. Even before simulations are performed, it is possible to make some predictions using concepts from polymer physics – the key idea is that if a polymer is represented as a chain of connected units which are otherwise free to move in space, it then looks like a random walk (or a self-avoiding walk) which has well known statistics[7]. Thus, questions such as "*On average how much space will a polymer of given length take up?*" or "*What is the probability that the polymer will be forming a loop?*" can be addressed. Complexity can be subsequently added to the description, for example by considering the flexibility of a polymer, or the solvent conditions. These statistical methods have been applied, for instance, to predict how the average level of Hi-C interactions decays with genomic separation[19], or how likely chromatin loops of certain length are to form. Often it is when an observed quantity *does not* fit the statistical description that we learn something – if the statistics of a free polymer predict one result, but something else is observed (either experimentally or computationally), there must be some mechanism we have not considered.

MD simulations and their variants are mature methods in the physics community and many software packages are available. The "bead-chain" simulations described below are often performed using general-purpose and adaptable open-source MD codes. The advantages of such software include their computational efficiency, their scalability when run on computer clusters, and their broad user base who keep software maintained and provide updates. In Box 2, we list some of the most popular software packages, and outline how they are used and on what type of computer hardware.

**Evolution and success of polymer models**

Recent mechanistic models, which treat chromatin as a simple polymer, have improved the interpretation of results from Hi-C and other experiments.

To study chromatin at a genomic scale, some of the first polymer simulations considered the role of entropic effects in chromosome positioning, showing that chromatin fibre properties such as flexibility can control radial positioning of chromosomes or chromosome regions[20]. Another study[21] used simulations and polymer theory to estimate time-scales for chromosome dynamics, revealing that the time required for two human chromosomes to become intermingled is of the order of 100s of years, which reconciles the observations that individual chromatin loci can be highly dynamic (e.g. structural changes after an inflammatory response can be observed within as little as 30 mins), while whole chromosomes appear relatively static (remaining in territories for the duration of the cell cycle[21]).

A more recent model considered the formation of chromosome domains. Known as the "strings-and-binders-switch" model[22], it combined a bead-chain polymer with generalized "binders": single beads representing chromatin-binding protein complexes that can form bridges between different chromatin regions. In this model, the bridging binders can generate chromatin interactions with the right scaling as a function of genomic separation, and provide a mechanistic understanding of the observation that there are different levels of compaction in different chromatin regions. At around the same time, we also developed a bead-chain model with diffusing bridges (or "factors")[23], and performed simulations of large chromosome regions (up to whole chromosomes) at a higher resolution (one to three kbps of DNA per bead). We uncovered a mechanism called the "bridging-induced attraction" – see Figure 2A. Put simply, protein complexes which can form molecular bridges between multiple regions of DNA or chromatin (stabilising loops) have a tendency to cluster together, even in the absence of interactions between the complexes. The attraction arises due to a positive feedback where the first formed loop creates a local increase in DNA density which promotes the binding of further proteins in that region. This process further increases DNA density and provides a positive feedback. The resulting cluster formation can be thought of as phase separation (or, more precisely, micro phase separation, as multiple clusters remain in steady state) – a mechanism now thought to be important for the formation of many nuclear structures[24]. Later work showed that the modelled polymer can be driven into specific 3-D structures[25] by including multiple species of bridges/factors and patterning the bead-chain with binding sites for these factors. It is possible to predict Hi-C maps by using experimental data to infer the positions of the binding sites (Figure 2A top right; and since this is a forward modelling scheme, it is truly a prediction as Hi-C data was not an input to the model). Rather than representing specific known species of protein or protein complex, the factors are assumed to be generic chromatin binders; and then rather than identifying precise binding sites, data such as ChIP-seq for different histone modifications are used to identify broad regions of binding. With this "simplistic" model[25] it was found that remarkably it was possible to reproduce the interactions *vs* genomic separation scaling and predict the locations of 85% of chromatin domain boundaries with only two bridge species: an active factor (e.g., polymerase/transcription factor complexes), and a repressive factor (e.g., polycomb complexes, or heterochromatin protein 1).

These "transcription factor" (or "diffusing bridge") models also provide predictions on the spatial organisation and dynamics of bridge complexes. The protein clusters which arise in the simulations look similar to some of the phase separated membrane-less organelles. For example, clusters of polycomb like proteins resemble polycomb bodies, and clusters of activating proteins can be thought of as transcription factories[30]. However, a more detailed inspection of the simulations showed that the dynamics of the protein clusters are not the same as nuclear bodies in an important respect: once formed, the clusters are very stable and proteins are not "turned over" (exchanged with a soluble pool). A refined model[31] adds a feature where the proteins stochastically switch from a binding to a non-binding state. This could represent post-translational modifications, active protein degradation, or programmed polymerase unbinding after transcription termination. The refined model leads to more dynamic nuclear body-like clusters where proteins can exchange with those in the soluble pool while the body retains its shape and size (Figure 2A, bottom right). Importantly the switching between protein states drives the system away from equilibrium (it represents active chemical reactions which hydrolyse ATP) and provides a mechanism through which the cell can control protein clustering and concomitantly phase separation. The switching model is an example where a discrepancy between the original simulations and the experimental observations led to model refinement and thus improved understanding.

Active protein unbinding is just one example of an out-of-equilibrium process which affects genome organisation. Biological systems are inherently far from equilibrium, as they take in energy from their surroundings to drive internal processes. Out-of-equilibrium processes are therefore often important. A challenge in coarse-grained modelling is to recognise when including such processes is necessary to explain the behaviour of the system. Active processes break "detailed balance", which can dramatically affect macroscopic behaviour. At equilibrium, as each process in the system must be balanced by its reverse process, a movie of a simulation looks essentially the same when played forwards or backwards: the time-reversal symmetry is typically lost when models include active out-of-equilibrium processes. Another important consideration is that as the cell exits mitosis, the compacted mitotic chromosomes expand into their interphase configurations – they relax towards a new equilibrium state but might not reach that state within biologically relevant time scales. This means that when running a simulation without an active process, one must consider carefully whether an equilibrium condition has been reached, whether such a state is relevant, and whether the outcome of the simulations will be affected by the initial configuration.

Another recent work[32] showed that to reproduce many of the features of a Hi-C map, it is not necessary to explicitly model the diffusing proteins – instead a direct attractive interaction between the polymer beads is sufficient. The direct attractive interaction could represent either bridging factors which are already bound to the chromatin, or direct chromatin-chromatin interactions mediated by charges on the nucleosome surface or histone tails. While those models (usually known as "block copolymer models") can give good prediction of Hi-C maps (for example in *Drosophila*), obviously they cannot give any information about protein foci dynamics without an explicit representation of the bridging proteins.

Another well-known model based on ideas developed through polymer physics is "loop extrusion"[26,27]. The loop extrusion model was recently used to explain a puzzling observation involving long-range interactions between binding sites of the CCCTC-binding factor (CTCF). This protein has a binding motif with a specific direction on the DNA, and interaction between binding sites tend only to be found when the binding motifs have a specific "convergent" orientation[4]. Such a strong bias is difficult to reconcile with loops that form due to two sites diffusing into contact. Instead, the model proposes that a factor binds at some point between the CTCF sites and extrudes the loop outwards (Figure 2B). There is growing evidence that the SMC protein cohesin is involved in this extrusion process, but it is still unclear if a unidirectional "motor" is required to push the chromatin into loops[26,27], or if diffusive sliding of cohesin is sufficient[28]. Nevertheless, the loop extrusion concept has been successful in providing clear qualitative explanations for experimental observations. It explains changes in chromatin interactions resulted from "genome editing" experiments where CTCF binding sites were manipulated[26]; it also explains changes observed in Hi-C data when CTCF[33], cohesin[34,35], or factors involved in cohesin loading and unloading[36,37] were knocked out. It is also thought that an extrusion mechanism, involving condensin, is involved in chromosome compaction during mitosis[29,38].

It is now believed that chromosome organisation is driven *both* by bridging protein complexes and some form of extrusion[39]. While extrusion can explain CTCF looping, it does not provide a mechanism for compartment formation; conversely, a bridging mechanism readily explains compartments but cannot generate CTCF loops with a motif direction bias. Taken together, these two physics-based models explain many of the observations from Hi-C experiments, including compartments, domains and loops. Particularly, the fact that compartment patterns remain, but loop-mediated TADs are lost from Hi-C maps when cohesin (or its loading factor) is removed suggests that these features arise through different mechanisms[34,35].

**Chromosome folding at the gene-scale**

The polymer models described above have mainly been applied to chromosome organisation at large scales, from hundreds of kilo-base to mega-base. Other efforts have focussed on an intermediate scale, studying the looping and folding of chromatin around specific genes and gene loci, which consider cis-regulatory promoter/enhancer interactions explicitly. Experimentally, higher resolution chromosome population level interaction data can be obtained using methods such as 4C or Capture-C[40], and single cell information can be collected through microscopy techniques such as fluorescent in-situ hybridisation (FISH). In simulations, greater levels of detail can be probed using different coarse-graining, e.g., by using more "beads" to represent the same length of chromatin.

Our work simulating looping of mouse globin genes[41] revealed that it is necessary to use different input data for the polymer model to capture higher-resolution interactions. Although using histone modification data to infer protein binding can give good predictions of domains and compartments, more specific binding site placement is required to predict promoter-enhancer contacts. For example, ChIP data for transcription factors can be used, but we observed that Capture-C results could be recapitulated using only DNA accessibility information derived from DNase- or ATAC-seq experiments. Importantly, these protein-binding driven simulations could generate an ensemble of locus conformations which were largely consistent with both population (Capture-C) and single cell (FISH) data. An important feature of these simulations is that the full details of each locus conformation within the population are retained, allowing the variability of the locus structure to be examined in a way that is still not possible experimentally. For the globin genes, analysis of the simulations implied that these loci tend to organise into one of a small number of possible structures, for example, forming a single compacted globule, two separated globules, or more extended shapes.

We applied a similar model – but with incorporation of diffusing bridges *and* loop extruders – to the developmental gene *Pax6*, in cells where the gene has different levels of transcriptional activity revealed different behaviour[42]. First, this model gave good predictions of interactions at the population level (simulating Capture-C data), but it failed to correctly predict the shape of the locus at the single cell level (simulating FISH data). Experimentally when *Pax6* was transcribed at a low to moderate level, the locus was more compact compared to the case where the gene was inactive (the separation between the *Pax6* promoter and its enhancers decreased). Surprisingly in a cell line where *Pax6* was highly active, the locus became more expanded with separations between the promoter and one enhancer significantly increasing. This variation was not correctly captured by the simulations; again, and importantly, the failure of the model led to new ideas for how the chromatin within the locus changes in these different cell types. The data showed that although large regions of the locus gained a histone acetylation mark associated with active enhancers (H3K27ac), this process was not accompanied by an increase in looping between those regions and the promoters as expected from the classic model of enhancer action. We reasoned that the mark was instead associated with some local change in the properties of chromatin which led us to a new "heteromorphic polymer" or HiP-HoP model[42] (Figure 3A). We hypothesised that the acetylation mark corresponded to regions of the chromatin with a less compact internal structure. This idea was previously suggested by experiments such as RICC-seq[43]. Indeed a simulation model which incorporates a fibre with varying linear compaction (i.e., different regions have a different amount of DNA per unit length) was able to reproduce all of the trends observed in the Capture-C and FISH data (Figure 3B-C). HiP-HoP can now be used to analyse the structure of many other genes across different species (e.g. *SOX2*), with the aim of improving our understanding of the link between structure and function. For example, the simulations of *Pax6* revealed a much larger variation of structure within a population of the same cell type compared to the case of the globin genes, suggesting that regulation of different genes involves alternate structural mechanisms.

An important point which arose from the *Pax6* study is that the classic model for enhancer-promoter interaction does not always seem to be at work. When comparing the inactive and low expression states, the distal *Pax6* enhancers gain histone acetylation marks and physical interactions with the promoters– as expected. Moving to the high expression cell line, for one enhancer the acetylated region broadens, but the frequency of interactions with the promoters *decreases* (and their separation *increases*); either this site does not have an enhancer action in these cells, or it functions by a mechanism other than physical contact with the promoter. The action is different for different expression states. On a technical level, we note that the simple nature of these models allows the large-scale behaviour of the chromatin to be simulated, without needing a detailed knowledge of the molecular details, so questions such as, is it a 30-nm or 10-nm fibre, a one-start or two-start helix, do not have to be addressed.

**Outlook**

There is a growing body of work using mechanistic polymer models to understand chromosome organisation and function. Importantly, it is clear that these methods are not only useful when applied to existing experimental data, but that simulations can also be used to test new ideas, to uncover new mechanisms, and to drive new experimental studies. A good example is the recent work on loop extrusion: the initial theoretical endeavour [27] rapidly prompted further experimental studies[26,33-37]. In physics it is commonplace that computational work is conceptually ahead of what can be realised experimentally, and we believe that in the future this will become more prevalent in chromatin biology.

What does the future hold for polymer simulations in chromatin biology? Our recent HiP-HoP model suggests that if we want to study gene loci in more detail, we need models that resolve some of the structural properties of chromatin and how they vary at different locations. There are models which have detailed coarse-grained representations of nucleosomes[16,18]– this approach has recently been used to study the *HOXC* locus[44], and revealed that epigenetic factors play an important role in larger-scale locus folding. But these detailed simulations are computationally expensive, and tend to be limited to small fibre sections – there is currently no model which resolves nucleosomes simply enough to allow simulation of large gene loci. If such a model were developed, it could for example be used to study how specific patterns of nucleosome spacing around regulatory elements affect the 3-D fibre structure, or to better understand data resulting from new experimental techniques such as ChromEMT[45] (where electron microscopy is used to image nucleosomes *in vivo*).

So, does it always benefit to have models with more molecular detail? Not necessarily – the level of detail required in a model really depends on the questions to be asked, and the size of the system being investigated (see Table 1). For example, recent studies using less detailed models to examine compartmentalization and the global organisation of (hetero)chromatin within the nucleus have revealed that interactions between the chromatin and the nuclear lamina play a key role, e.g. in the inverted structure observed in rod cells in nocturnal mammals[46], and in structural changes during senescence and ageing related diseases[47]. Similar low-resolution models might be appropriate to study how chromosomes are compacted during mitosis[38], or to study the kinetics of chromatid segregation during anaphase. Models with different scales and levels of detail might also be useful for gaining a general understanding of mechanisms or phenomena such as DNA supercoiling, liquid-liquid phase separation of chromatin associated proteins, or active processes which occur within the nucleus.

Polymer and coarse-grained or "mesoscale" simulations have proven useful for studying soft matter physics. We expect such methods become useful tools for chromatin biology. Challenges remain in developing models which incorporate pertinent molecular detail at a short length scale, while addressing large-scale behaviour and simulating physiologically relevant time scales. These exciting new applications for mesoscale modelling are therefore likely to drive new developments in multi-scale simulation methods, where systems are simultaneously modelled with different levels of detail, with results from one model feeding into another.

## Acknowledgements

The authors would like to thank members of their groups for stimulating discussions. Research in the Marenduzzo group is supported by the European Research Council (CoG 648050, THREEDCELLPHYSICS); research in the Gilbert lab is funded by the UK Medical Research Council (MR/J00913X/1 and MC_UU_00007/13).

## Author Contributions

CAB designed and co-wrote the manuscript. DM co-wrote the manuscript. GN conceived and co-wrote the manuscript.

## Ethics Declaration

The authors declare no competing interests.

**Figure 1:** Coarse-grained molecular dynamics (MD) simulations of chromatin. **A.** Schematic description of classic MD. - **B.** A simplified description of the system reduces computational overhead. (i) Solvent atoms (red and grey) are removed, and their effects are provided implicitly by adding random force "kicks" to the solute atoms (represented by arrows, also see Box 1). (ii) Coarse-graining is where collections of atoms are replaced by simpler objects. Here a coarse-grained model represents each nucleotide as a bead with a patch[14]. (iii) Chromatin can be coarse-grained at different levels depending on the question being asked (see Table 1). From left to right: an all-atom representation of two nucleosomes (based on the crystal structure from Schalch et al.[15]), a chromatin fibre model where each nucleosome is represented by a solid body (for a review of such models see Schlick et al.[16]), and a bead-chain model where the internal structure of the fibre is not resolved.

**Figure 2:** Commonly used models for understanding chromosome organisation. **A.** Left: Diffusing protein bridges stabilize loops and domains in chromatin or DNA, leading to bridging induced attraction (even in the absence of protein-protein interactions). Right top: specific patterns of binding sites on chromatin lead to domains and compartments (predicting Hi-C maps[25] from protein binding data). Right bottom: ATP-driven chemical reactions altering bridging affinity lead to clusters with dynamics similar to nuclear bodies. **B.** The loop extrusion model[26,27] explains chromatin loop domains and the CTCF motif directionality bias. Extrusion could be an active "motor" effect, or diffusive[28] and a similar mechanism might drive chromosome compaction during mitosis[29].

**Figure 3:** A gene locus model: highly predictive heteromorphic polymer model (HiP-HoP)[42]. **A.** Schematic showing the HiP-HoP model ingredients – a variable thickness fibre is combined with diffusing bridges and loop extrusion. **B-C.** This model was used to study the *Pax6* locus in three cell lines where *Pax6* is expressed at different levels. The predictions are confirmed by population level chromatin interactions (Capture-C) and fluorescence in situ hybridization. In B, coloured lines show simulation results and grey bars show experimental data. In C, the scale bar in the image is 0.5 $\mu$m, and the box plots show separations of probes located at Pax6 and a downstream enhancer..

***Box 1: Langevin dynamics for modelling molecular motion***

Running a molecular dynamics simulation essentially boils down to solving Newton's second law (the "equation of motion" or $F = ma$) for each atom (or coarse-grained (CG) object) in the system. To simplify the simulations, the full details of the solvent can be neglected – a common scheme for doing this is Langevin dynamics, in which the effect of the solvent is approximated by two additional forces to the equation of motion. Here we look at this in more detail. The equation of motion for atom $i$ is

$$m\frac{d^2\boldsymbol{r}_i}{dt^2} = \boldsymbol{F}_i - \xi\frac{d\boldsymbol{r}_i}{dt} + \sqrt{6\xi k_B T}\boldsymbol{\eta}_i(t),$$

where $\boldsymbol{r}_i$ is the vector position in space of atom $i$. The left-hand side of the equation is mass times acceleration (the second derivative of position with respect to time). The three terms on the right are the forces experienced by atom $i$. In a simulation, this equation is solved numerically – by imagining that time evolves in discrete steps, we calculate the forces on the atom at one time point, and this equation tells us how those lead to a change in the velocity and position of the atom at the next time point. Each of the simulation software packages mentioned in the main text essentially solves this equation for all of the atoms (or CG objects) in the system.

The three force terms:

$\boldsymbol{F}_i$      This is the force experienced by "atom" (or bead) $i$ due to interactions with all other atoms in the system. For an atomistic simulation this is found from a complex set of interaction potentials derived from quantum chemistry; for a CG simulation it may be a set of simplified phenomenological potentials. In a typical CG polymer model, interactions could include a potential to keep beads connected in a chain, a potential preventing beads from overlapping in space, and a potential giving rise to a polymer bending stiffness. This term could also include any external force applied on bead $i$.

$-\xi\frac{d\boldsymbol{r}_i}{dt}$      This is the first of two terms which approximate the effects of the solvent. It represents the viscous drag experienced by the bead as it moves, which is proportional (and in the opposite direction) to its velocity; $\xi$ is a "friction" parameter related to the viscosity of the fluid.

$\sqrt{6\xi k_B T}\,\boldsymbol{\eta}_i(t)$ This second solvent term approximates thermal "jostling" caused by solvent molecule interactions. The symbol $\boldsymbol{\eta}_i(t)$ represents a random "kick" of force bead $i$ receives at time $t$.

There is a well-defined mathematical description of this "noise", but in a simulation context it amounts to generating a random number at each time step – which introduces stochasticity. The pre-factor ensures that the equation obeys the fluctuation dissipation theorem: in the context of Brownian motion there is a relationship between the viscous drag experienced by an object being pulled through a fluid (dissipation) and its diffusive motion (fluctuations).

This simple scheme neglects effects like hydrodynamics interactions (where fluid flows set up by motion of one object result in forces on another). Including these would require significant computational overhead, and generally it is thought that they should not play a big role in the densely packed nuclear environment.

***Box 2: Practicalities of Polymer Simulations***

As detailed in the main text, most polymer-based molecular dynamics (MD) simulations of chromatin are performed using general-purpose codes. Common software packages include: LAMMPS (Large-scale Atomic/Molecular Massively Parallel Simulator[48]) which is optimized for use on multicore computer clusters and is an adaptable and expandable code written in C++; HOOMD[49], which uses the Python scripting language and is optimised to run on GPUs; and ESPResSo (Extensible Simulation Package for Research on Soft Matter[50]), which also uses Python and was developed for soft matter physics. These packages come with extensive documentation, and tutorials are often a good starting point for new users (e.g., see cbrackley.github.io/simple_lammps_tutorial). The simulations output "trajectories", i.e. details of the positions of atoms (or CG objects) as a function of time; since an implicit solvent simulation is stochastic, many such simulations can be performed to generate an ensemble of trajectories (representing e.g. a population of cells). Typically, users write further programs or scripts to take measurements from these trajectories which are compared to experimental measurements (e.g. one might measure the separation of CG objects, diffusion constants, or how often two objects are found together within a population of trajectories). Other software tools, such as Visual Molecular Dynamics[51] (VMD), are used to visualise trajectories.

The software mentioned above can be compiled to run on any standard Unix-based system (e.g. Mac or Linux). Though most software can run on any type of machine, from laptop or desktop up to multi-core supercomputers, the scale of most studies necessitates the use of multi-core machines (or alternative architectures like GPUs or other coprocessors). A typical simulation-based study might need thousands of CPU hours-worth of compute time and generate GBs of data. This requirement for specialist high-performance computing hardware means that these simulations should be viewed as "*in silico* experiments" in that – just as a wet-lab based project – specialist equipment, expertise, technicians and consumables (here compute time and data storage) are required.

**Table 1.** Different coarse-grained polymer simulation models can be, or have been, used to study different biological questions. "Mechanistic" models are used to test hypothesis on underlying mechanisms; and "inverse" models, where data are used to, e.g., infer chromatin structures consistent with these.

| *Mechanistic (forward) models:* | |
|---|---|
| *Aim* | *Examples* |
| *Understand microscopic DNA properties, in vitro experiments, or biotechnology applications.* | These often involve CG models of DNA which resolve the DNA double strands and superhelical structure, e.g., the oxDNA model[17], or three-spheres-per-nucleotide (3SPN) model[52]. These can be used to study supercoiling[14], DNA melting, and DNA-protein interactions. Larger systems can be treated using simpler models which track twist deformations without resolving individual DNA strands[53]. |
| *Understand chromatin structure at the nucleosome level.* | Detailed CG models based on nucleosome crystal structures can be used to simulate small fibres[18,54]. Simplified models representing nucleosomes as disks or spheres have been used to simulate, e.g. in vitro nucleosome unwrapping[55], chromatin reconstitution[56], or micro-domains in yeast[57]. |
| *Understand mechanisms of chromosome compartments and domains.* | Bead-and-spring polymer models for chromatin (where nucleosomes are not resolved) can be used to investigate mechanisms for the formation of compartments and domains in, e.g. Drosophila[32], mouse[58], and humans[25]. |
| *Understand mechanisms for chromosome loops and loop domains.* | Coarse-grained bead-and-spring polymer models can be used to study these questions. On the basis of such studies, several different mechanisms have been put forward for the formation of cohesin and CTCF mediated chromatin loops, and loop domains. These include supercoiling[59], and loop extrusion[26–29]. |
| *Predict the detailed structure of gene loci* | Coarse-grained bead-and-spring polymer models can also be predictive, using some experimental data as an input. Unlike the inverse models detailed below, they do not involve fitting. The HiP-HoP model[41,42] is an example: data on DNA accessibility is an input, and simulated Hi-C/Capture-C is the output (see also ([60])). |
| *Understand chromosome organisation at the whole nucleus level* | By reducing the level of detail (e.g. representing large chromosome regions as a single bead), whole chromosomes[20,21], or even whole nuclei can be simulated. This has been used, e.g. to study the effect of interactions between chromatin and the nuclear lamina[46,47]. |
| *Inverse Models:* | | |
| *Aim* | *Examples* | *Input data* |
| *Generate in silico configurations of a gene locus, or whole chromosome consistent with 3C-based data.* | To reconstruct chromosome configurations from data, restraint-based and polymer-based models have been developed. Examples of the former include the TADbit software, which has been used to reconstruct gene loci in mammalian cells[12,61]; similar methods have been applied to reconstruct whole yeast nuclei[62]. In polymer-based methods, iterative parameter determination has been combined with both Monte Carlo[10] and Molecular Dynamics simulations[11,63] to generate structures from 5C and Hi- | *Hi-C, 5C or single cell Hi-C* |

| | C data. By reducing model resolution, whole human chromosomes, or even whole nuclei can be modeled, e.g. to reconstruct configurations from single cell Hi-C data[64]. | |
|---|---|---|
| *Predict the effect of genome rearrangements* | In some versions of the above models, once parameters are generated from one Hi-C data set, they can be used to make predictions about the effect of genome rearrangements[63], or cell differentiation[11]. | *Hi-C of the "wild type"* |

## A. Molecular Dynamics

Positions of atoms tracked

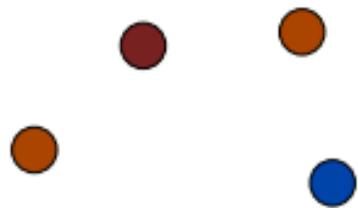

Atoms exert forces on each other

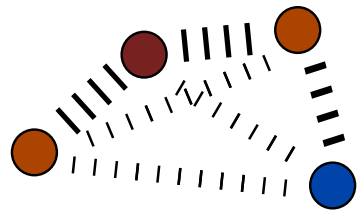

Motion simulated by solving Newton's Law

$$m\frac{d^2\mathbf{x}}{dt^2} = \mathbf{F}$$

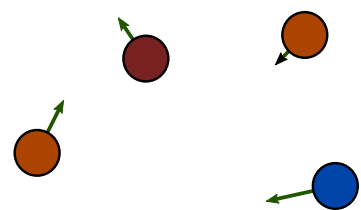

## B. Coarse Graining

(i) implicit solvent

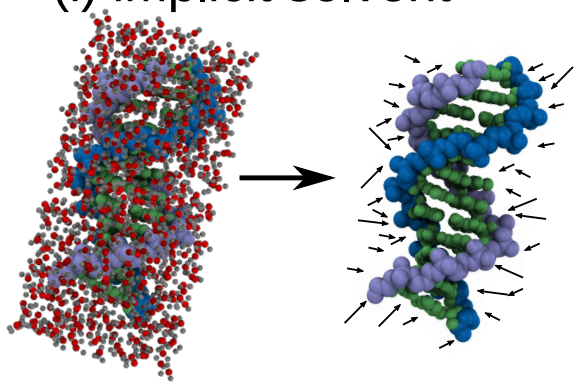

(ii) coarse-grained DNA

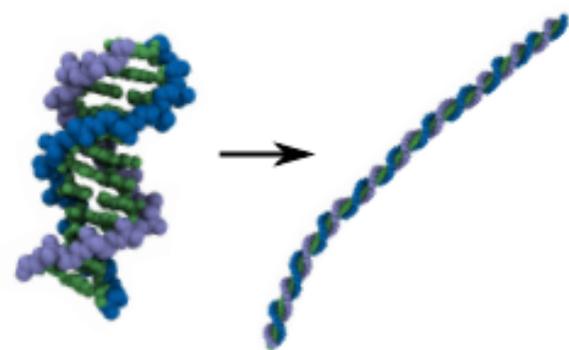

(iii) coarse-grained chromatin

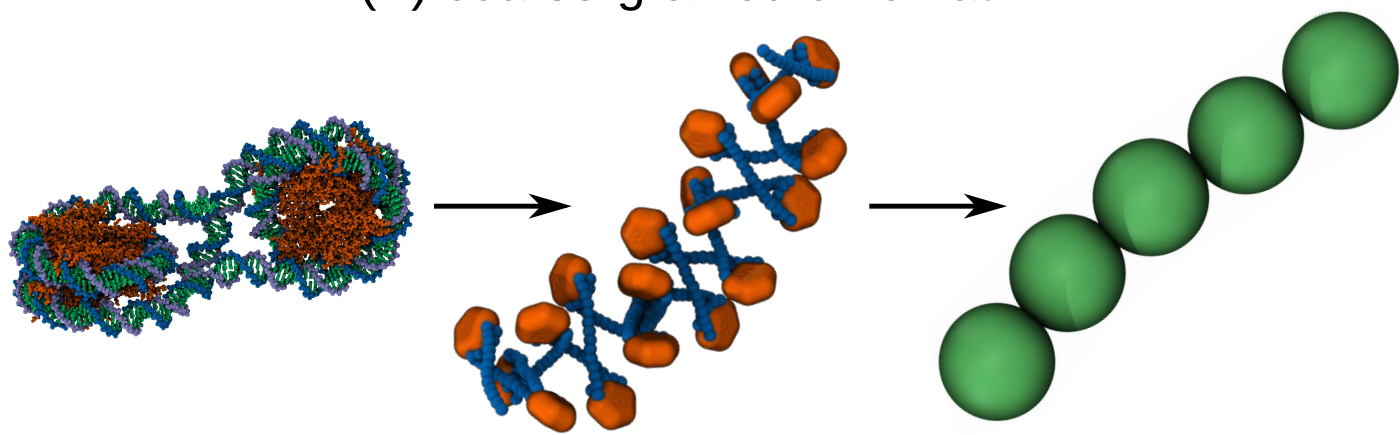

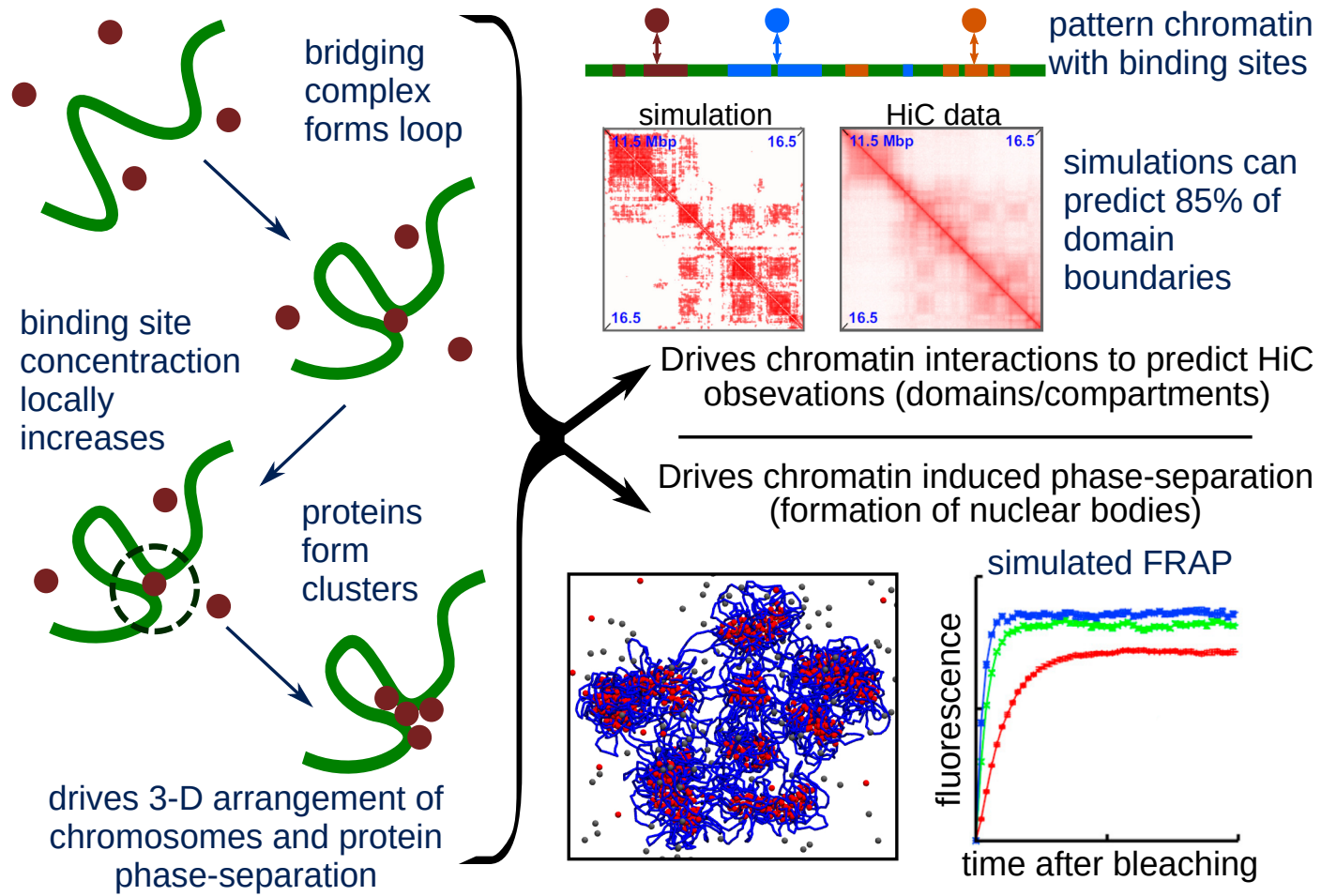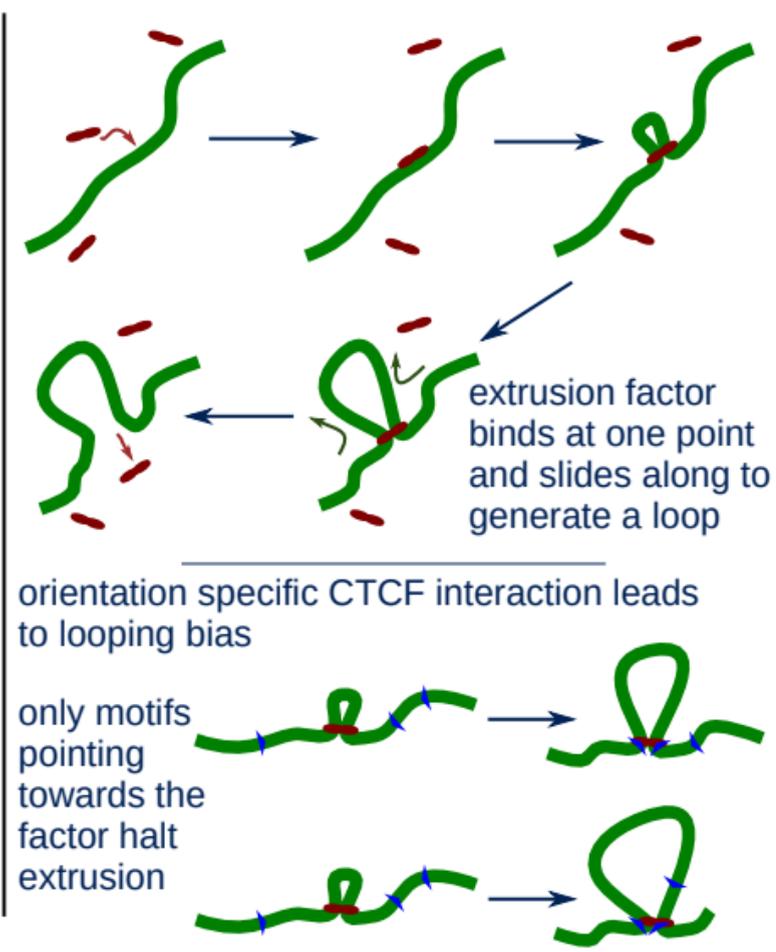

## A. HiP-HoP Model

**HI**ghly **P**redictive **H**eter**O**morphic **P**olymer

variable fibre thickness

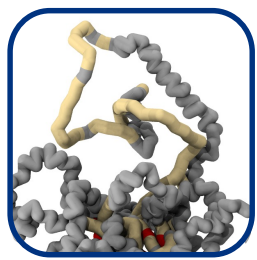
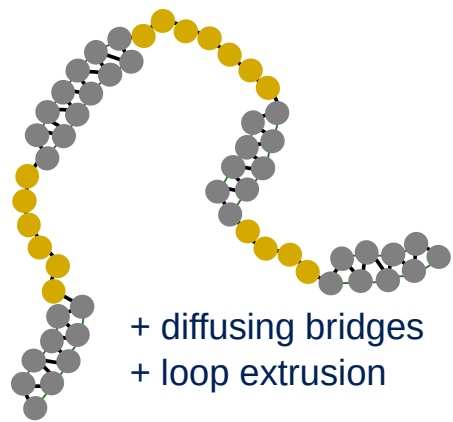

+ diffusing bridges
+ loop extrusion

Input data: *DNA accessibility (e.g. ATAC-seq)*
*CTCF / Cohesin ChIP-seq*
*H3K27ac ChIP-seq*

Output: *a population of locus conformations*

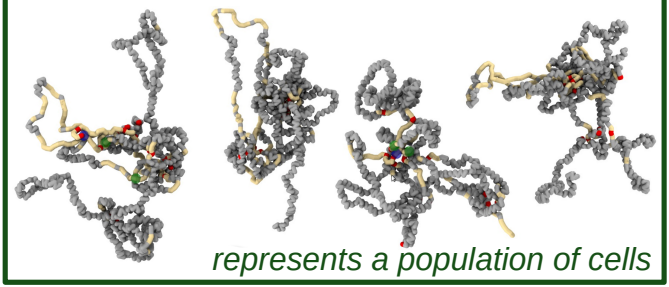

*represents a population of cells*

## B. 3C-style measurements at the Pax6 Locus

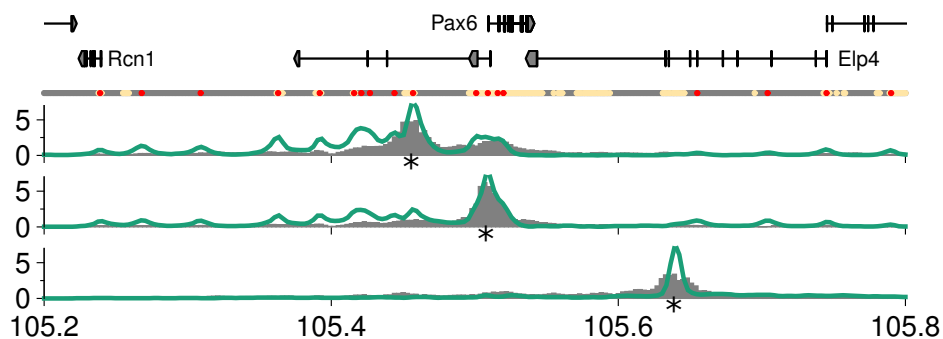

population level information can be compared to HiC / 4C / CaptureC etc.

the full set of conformations can be studied, e.g. to look for variation within a population

## C. FISH at the Pax6 Locus

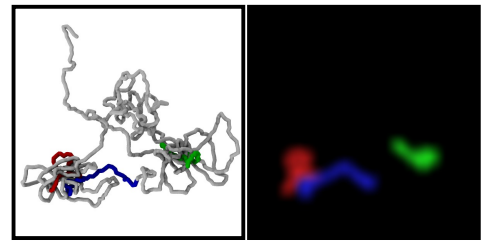

simulation

compare with experiment

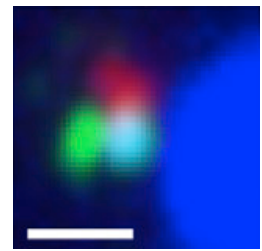

single cell-like information can be compared to microscopy

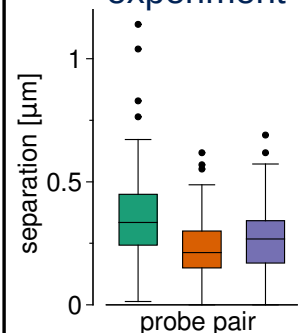
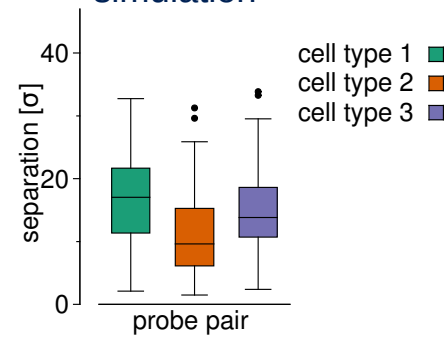

experiment — simulation

cell type 1
cell type 2
cell type 3